\documentclass[12pt]{article}
\usepackage{graphicx,cite,epsfig}
\newcommand{\ba}{\begin{array}}
\newcommand{\ea}{\end{array}}
\newcommand{\beq}{\begin{equation}}
\newcommand{\eeq}{\end{equation}}
\newcommand{\comment}[1]{}
\def\bt{\begin{table}}
\def\et{\end{table}}
\def\bc{\begin{center}}
\def\ec{\end{center}}
\def\bi{\begin{itemize}}
\def\ei{\end{itemize}}
\def\bea{\begin{eqnarray}}
\def\eea{\end{eqnarray}}
\def\beas{\begin{eqnarray*}}
\def\eeas{\end{eqnarray*}}

\def\N0{\widetilde{\chi}^0}

\def\snu{\tilde{\nu}}

\def\slep{\widetilde{\ell}}

\def\PpL{P^{e^+}_L}
\def\PeL{P^{e^-}_L}
\def\PeT{P^{e^-}_T}
\def\PpT{P^{e^+}_T}
\def\ee{e^+e^-}

\def\a{\alpha}

\def\g{\gamma}

\def\G{\Gamma}

\catcode`@=11 
\def \gsim{\mathrel{\mathpalette\@versim>}}
\def \lsim{\mathrel{\mathpalette\@versim<}}
\def \@versim#1#2{\lower0.4ex\vbox{\baselineskip\z@skip\lineskip\z@skip
     \lineskiplimit\z@\ialign{$\m@th#1\hfil##\hfil$%
     \crcr#2\crcr\sim\crcr}}}
\catcode`@=12 
\begin{document}

\begin{flushright} HIP-2009-04/TH \end{flushright}
\begin{center}
{\Large Use of Transverse polarization to probe  R-parity violating
supersymmetry at ILC.}
\\ \vspace*{0.2in}
{\large Rohini M Godbole$^1$, Santosh Kumar Rai$^2$ and Saurabh D Rindani$^3$} \\
\vspace*{0.2in}
{\sl $^1$Center for High Energy Physics, Indian Institute of Science,
         Bangalore 560 012, India\\
     $^2$ Department of Physics, University of Helsinki and
                                 Helsinki Institute of Physics, \\
          P.O. Box 64, FIN-00014, Helsinki, Finland\\
     $^3$ Theoretical Physics Division, Physical Research Laboratory,
          Navrangpura \\ Ahmedabad 380 009, India
\rm }

\end{center}
\vspace*{0.6in}
{\large\bf Abstract}

In supersymmetric theories with R-parity violation, squarks and sleptons can
mediate Standard Model fermion-fermion scattering processes. These scalar 
exchanges in $\ee$ initiated reactions can give new signals at future linear
colliders. We explore use of transverse beam polarization in the study of these signals  in the process  $e^+e^- \to b\bar b$. We highlight certain asymmetries,
which can be constructed due to the existence of the transverse beam 
polarization, which offer discrimination from the Standard Model (SM) background
and provide increased sensitivity to the R-parity violating couplings.


\rm\normalsize
\section{Introduction} \label{sec-1}
In the Standard Model (SM), baryon and lepton number conservation
is not guaranteed by local gauge invariance. In fact, in the
supersymmetric extension of the SM, 
the most general superpotential respecting the gauge symmetries of the
SM contains bilinear and trilinear terms which do not
conserve either of baryon ($B$) and lepton ($L$) numbers.
Clearly, the simultaneous presence of both lepton- and baryon-number
violating operators could lead to very rapid proton decay, especially
for TeV scale sparticle masses.
The existence of all such terms can be forbidden by postulating a
discrete symmetry~\cite{Fayet:1977yc},
called R-parity, which implies a conserved quantum number
$R_p \equiv (-1)^{3B + L +S}$, where $S$ stands for the spin of the
particle. The very definition implies that all the SM particles
have $R_p = +1$ while all the superpartners are odd under
this symmetry. Thus, apart from suppressing proton decay,
it also guarantees the stability
of the lightest supersymmetric particle (LSP) thereby
offering a ready-made candidate for cold dark matter.

However, while a conserved R-parity seems desirable,
it is perhaps too strong a requirement to be imposed.
For one, this symmetry is an {\em ad hoc} measure and
there does not exist an overriding
theoretical motivation for imposing it,
especially since a suppression of proton decay rate
could as well be achieved
by ensuring that one of $B$ and $L$ is conserved.
Indeed, it has been argued~\cite{Ibanez:1992pr} that
this goal is better served by imposing a generalized
baryon parity instead. Unlike R-parity, this latter
($Z_3$) symmetry
also serves to eliminate dimension-{\em five} operators that
could potentially have led to proton decay.
Furthermore, non-zero R-parity violating (RPV) couplings
provide a means of generating the small neutrino
masses, either at tree level or loop level,
that the neutrino oscillation experiments
seem to call for. It is thus of both theoretical and
phenomenological interest to consider violations of
R-parity~\cite{books}.

The most general R-parity violating superpotential is
\begin{eqnarray}
W \supset \sum_{i} \kappa_i L_i H_2 +
     \sum_{i,j,k} \bigg (\lambda_{ijk} L_iL_j E^c_k+
\lambda'_{ijk} L_i Q_j D^c_k+ \lambda''_{ijk} U_i^cD_j^cD_k^c
\bigg )
\label{super}
\end{eqnarray}
where $i,j,k$ are generation indices,
$L$ ($Q$) denote the left-handed lepton (quark) superfields,
and $E$, $D$ and $U$ respectively
are the right-handed superfields
for charged leptons, down and up-type quarks.
The couplings $\lambda_{ijk}$ and $\lambda''_{ijk}$ are antisymmetric
in the first two and the last two indices respectively.
A conserved baryon number requires that all the $\lambda''_{ijk}$
vanish identically thereby avoiding rapid proton decay.
Neutrino masses, being very small, restrict quite strongly the size of
the dimensional couplings $\kappa_i$ in Eq.~(\ref{super}) and of the vacuum
expectation values (\emph{vev}'s) of the neutral scalar components of the 
fields $L_i$, $v_i$. Note, however, that strictly speaking it is also  
possible to construct models with $\kappa_i, i=1,3$ not necessarily small. 
In this note,
we focus on the effect of the trilinear terms in the superpotential.

Written in terms of the component fields these terms
lead to the interaction Lagrangians
\beq
\ba{rcl}
{\cal L}_{LL\bar{E}} &
= & \frac{1}{2} \lambda_{ijk} \bigg[
   \snu_{iL}     \bar{\ell}_{kR}         \ell_{jL}
+  \slep_{jL}    \bar{\ell}_{kR}         \nu_{iL}
+ (\slep_{kR})^* \overline{(\nu_{iL})^c} \ell_{jL}
-  ( i \leftrightarrow j)
\bigg]  + h.c. \,
\ea
     \label{gen_Lag1}
\eeq
and
\beq
\ba{rcl}
{\cal L}_{LQ\bar{D}} &
= & \lambda'_{ijk} \bigg[
   \snu_{iL}     \bar{d}_{kR}         d_{jL}
+  \widetilde{d}_{jL}    \bar{d}_{kR}         \nu_{iL}
+ (\widetilde{d}_{kR})^* \overline{(\nu_{iL})^c} d_{jL} \bigg. \\
&& \bigg. -  \slep_{iL}            \bar{d}_{kR}            u_{jL}
- \widetilde{u}_{jL}     \bar{d}_{kR}          \ell_{iL}
- (\widetilde{d}_{kR})^* \overline{(\ell_{iL})^c} u_{jL}
\bigg] + ~ h.c.
\ea
     \label{gen_Lag2}
\eeq

\noindent
Just like the usual Yukawa couplings, the magnitudes of the couplings
$\lambda_{ijk},\lambda'_{ijk}$ are entirely arbitrary, and are restricted only from
phenomenological considerations.  The preservation of a GUT-generated
$B-L$ asymmetry, for example, necessitates the preservation of at
least one of the individual lepton numbers over cosmological time
scales~\cite{Dreiner:1992vm}.  Nonzero RPV couplings
mean a decaying LSP, whose decay may or may not be always prompt, 
and which is mostly taken to be a  neutralino~\cite{decay}, even though
non-$\tilde \chi_1^0, \tilde \tau_1$ candidates for the LSP are also
possible~\cite{dreinergrab}.
In all the cases the decaying LSP gives rise to striking collider
signatures~\cite{rpvcol}. However,  the failure so far of the various collider
experiments~\cite{nosusy,HERA} to find any evidence of supersymmetry
implies constraints on the parameter space.  Even if superpartners
are too heavy to be produced directly, their effects can still be probed
using low-energy observables~\cite{Barger:1989rk,books}.
The remarkable agreement of the measured values with the SM predictions
implies strong bounds on these couplings which
generally scale with the sfermion mass ($m_{\tilde f}$) \cite{books,rplimits}.
In this work we study processes directly sensitive to the size of such
couplings through the modification of SM amplitudes due to
sparticles exchange~\cite{Bhattacharyya:1994yc,
Hikasa:1999wy,CCQR}. The exchange
of spin-0 particles in a $2\to2$ scattering process would give a
completely different chiral behaviour to the amplitudes as compared to
the vectorial exchanges in the SM. The cleanliness of the
signal at the next generation International Linear Collider (ILC)~\cite{ILC}
and excellent reconstruction of the angular variables would help us
study the chiral properties of the amplitudes. The aim of this work is
to investigate use of transverse beam polarization to probe
such contributions  through the measurement of  cross-sections and
study of kinematical properties of the final states. Specifically, we
will see that transverse polarization can probe interference between
SM amplitudes and certain RPV mediated amplitudes
which are absent with longitudinally polarized or unpolarized beams.
As a result, the additional effects can depend quadratically on
the RPV couplings rather than quartically. This can make
studies with transversely polarized beams more sensitive to R-parity
violating couplings.

We concentrate on the simplest
process $\ee \to f\bar{f}$ at the ILC. We discuss the advantages of having
transversely polarized beams at ILC in Section \ref{sec-2} and its role
in addressing issues pertaining to the chiral nature of interactions. In
Section \ref{sec-3} we present the analysis and numerical results for the
process $\ee \to f\bar{f}$ and finally summarize and conclude in
Section \ref{sec-4}.

\section{Transverse polarization at ILC} \label{sec-2}
An $\ee$ linear collider operating at a center-of-mass energy of several
hundred GeV will offer an opportunity to make precision measurement of the
properties of the electroweak gauge bosons, top quarks, Higgs bosons, and also
to constrain new physics~\cite{ILC}.
Linear colliders are expected to  have the option of
longitudinally polarized beams, which could add to the sensitivity of
these measurements and reduce background in the search for new physics
\cite{gudi}. It has further
been realized that spin rotators can be used to convert the longitudinal beam
polarization to transverse polarization. This has inspired studies which
investigate the role of transverse polarization in constraining new
physics \cite{gudi,newphys,lepto},
though these studies are yet far from being exhaustive.

It was pointed out long ago by Hikasa \cite{hikasa}
that transverse polarization
can play a unique role in isolating chirality-violating couplings, to which
processes with longitudinally polarized beams are not sensitive.
This has been demonstrated recently in different situations \cite{lepto,ttbar}.

Polarization effects are different for chirality-conserving and
chirality-violating new interactions.  In the limit of vanishing electron
mass, there is no interference of the chirality-violating new interactions
with the chirality-conserving SM interactions.  As a result, in this
limit, any  contribution from chirality-violating interactions
which is polarization independent or dependent on longitudinal
polarization also vanishes.

Transverse polarization effects for the two cases are also different.
The cross terms of the SM amplitude with the amplitude from
chirality-conserving interactions has a part independent of transverse
polarization and a part which is bilinear in transverse polarization of
the electron and positron, denoted by $P_T^{e^-}$ and
$P_{T}^{e^+}$ respectively. For the case of chirality-violating
interactions, the cross term has only terms linear in $P_T^{e^-}$ and
$P_{T}^{e^+}$, and no contributions independent of these.

The interference of new chirality-violating contributions with the
chirality-conserving SM couplings give rise to terms in the
angular distribution proportional to $\sin\theta\cos\phi$ and
$\sin\theta\sin\phi$, where $\theta$ and $\phi$ are the
polar and azimuthal angles of a final-state particle. Chirality-conserving new
couplings, on the other hand, produce interference contributions
proportional to
$\sin^2\theta\,\cos2\phi$ and $\sin^2\theta\,\sin2\phi$. Chirality-violating
contributions do not interfere with the chirality-conserving SM
contribution with
unpolarized or longitudinally polarized beams when the electron mass is
neglected. Hence transverse polarization would enable measurement of
chirality-violating couplings through the azimuthal distributions.

In what follows, we will study a process which has an $s$-channel
contribution from scalars which violates chirality, as well as a
$t$-channel contribution from scalars which conserves chirality. In view
of the above remarks, the effects of these two kinds of contributions
will be different, and it is possible to study these separately.

\section { The process $\ee \to f \bar{f}$ at ILC} \label{sec-3}
It is needless to say that ILC will have the ability to make precise tests
of the structure of electroweak interactions at very short distances.
Looking at the simplest process of $e^+e^- \to f\bar{f}$, the SM
cross-section prediction can be put in the form

\bea
\frac{d\sigma (e^-_Le^+_R \to f_L\bar{f}_R)}{d\cos\theta}&=&
\frac{\pi\a^2}{2s}N_C
(1 + cos\theta)^2
\\ \nonumber
&\times&\left|Q_f + \frac{(\frac{1}{2}-\sin^2\theta_w)(T^3_f-Q_f\sin^2\theta_w)}
{\cos^2\theta_w\sin^2\theta_w}\frac{s}{s-m^2_Z}\right|^2
\eea
where $N_C=1$ for leptons and 3 times for quarks, $T^3_f$ is the weak
isospin of $f_L$, and $Q_f$ is the electric charge. For $f_L$ production,
the $Z$ contribution typically interferes with the photon constructively
for an $e^-_L$ beam and destructively for an $e^-_R$ beam. Thus,
initial-state polarization is a useful diagnostic at the ILC. Applied to
familiar particles, they would provide a diagnostic of the electroweak
exchanges that might reveal new heavy weak bosons or other types of new
interactions. We focus on the case when the beams are transversely polarized
and look at some specific processes which would be sensitive to couplings
as discussed in the previous section.
\subsection {Polarization study of $\ee \to b \bar{b}$}
We consider the process
\beq\label{process}
e^-(k_1,s_1)+e^+(k_2,s_2) \to b(p_1) + \bar{b}(p_2).
\eeq
In SM, this process proceeds via the $s$-channel exchange of $\gamma$
and $Z$. On including R-parity violation, the
process can receive contributions from
RPV couplings from the s-channel sneutrino exchange and the t-channel
sfermion exchange. The representative Feynman graphs for these latter
contributions are shown in Fig. 1(a) and
Fig. 1(b) respectively. The $s$--channel diagrams, for example,
involve chirality-conserving couplings
for the  exchange of a photon and a $Z$ and chirality-violating
couplings in an $s$-channel exchange of a sneutrino. In the absence of
any beam polarization, or with just the
longitudinal polarization, these two contributions do not interfere, and
the RPV couplings appear only at quartic order. With transverse
polarization, the interference between the vector and scalar exchanges
survive, giving rise to characteristic azimuthal distributions of the
type $\cos\phi$ and $\sin\phi$, which
enable discrimination of the RPV contribution from the SM
contribution, whose azimuthal dependence has the form $\cos 2\phi$ and
$\sin 2\phi$.
Since this contribution is at quadratic order in the RPV couplings,
transverse polarization leads to enhanced sensitivity to these
couplings. However, in case of the $b \bar{b}$ final state we consider,
the enhancement is unfortunately annulled by the suppression factor of $M_b^2/s$
arising because of the chirality-violating coupling of $b\bar b$ to the sneutrino.
The characteristically different azimuthal distribution because of the spin-0 sneutrino,
does, however, survive.

The $t$-channel sfermion exchange diagrams involving RPV couplings,
on the other hand, do interfere with the SM diagrams with longitudinal
or no polarization of $e^+$ and $e^-$. With transverse
polarization, they
give rise to azimuthal distributions of the same kind
as the pure SM contributions (with terms proportional to $\cos
2\phi$ and $\sin 2\phi$. However, their contributions to the
azimuthal distributions being quadratic rather than
quartic in the RPV couplings, they still offer a sensitive test of these
couplings.

We first write down the various terms contributing to the transition
probability for the process, and then study separately the contributions
of the RPV $s$-channel and $t$-channel exchanges.
\begin{figure}[htb]
\begin{center}
\includegraphics[height=2.0in,width=5.0in]{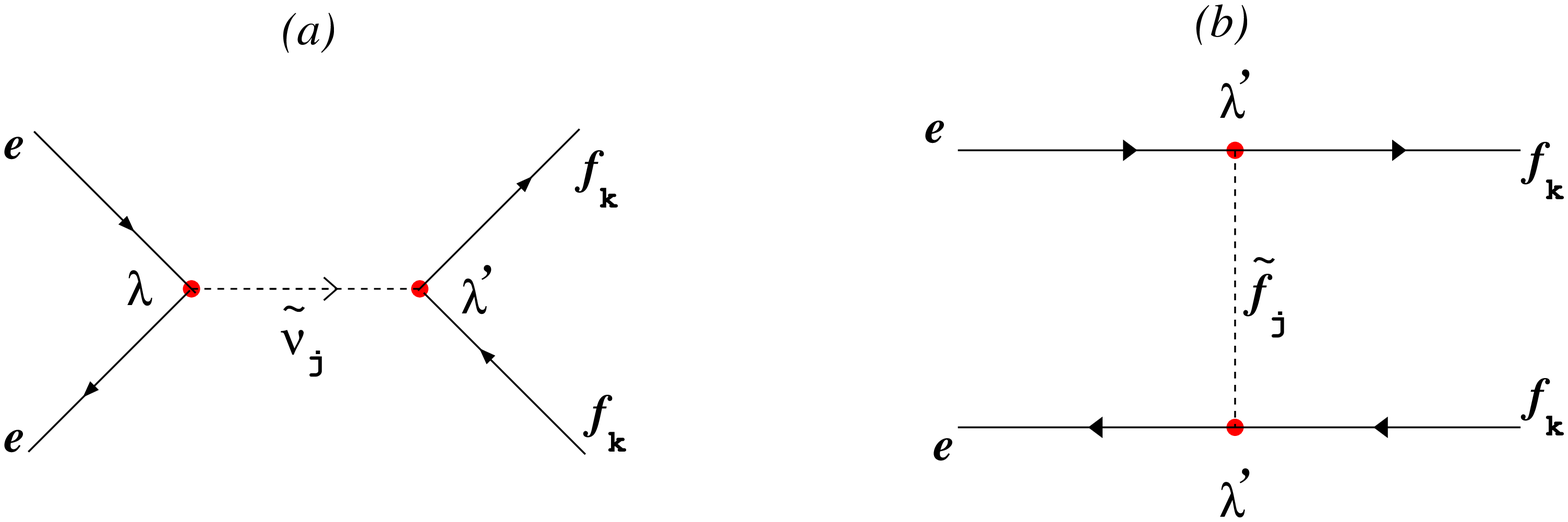}
\caption{\sl\small Feynman graphs for the RPV contributions
to the process $e^+e^- \to f_k\overline f_k$, where $f$ stands for any
fermion in the SM and $j,k$ represent the generation index.}
\label{fig1}
\end{center}
\end{figure}
We choose the following notation for introducing the beam polarizations
through the projection operators for electrons and positrons:
%
%
\beq \label{polsum}
\begin{array}{rcl}
\displaystyle
\sum_{s_1} u(k_1,s_1) \bar u(k_1,s_1)
= \frac{1}{2}(1 + P_L^{e^-} \g_5 + \g_5 P_T^{e^-}
t\!\!/_1) k\!\!\!/_1, \\
\displaystyle
\sum_{s_2} v(k_2,s_2) \bar v(k_2,s_2)
= \frac{1}{2}(1 - P_L^{e^+} \g_5 + \g_5 P_T^{e^+} t\!\!/_2 ) k\!\!\!/_2,
\end{array}
\eeq
where $t_{1,2}$ are the transverse polarization 4-vectors for the
electron and positron beams, respectively. In the above equation, $P_L$ and
$P_T$ represent the degrees of longitudinal and transverse
polarizations. For our analysis, we chose
$|P_{T}^{e^-}|=0.8$ and $|P_{T}^{e^+}|=0.6$. For the transverse
beam polarization 4-vectors we assume $t_1^\mu=(0,1,0,0)=-t_2^\mu$.

The process of Eq. \ref{process} is mediated by the $\g$ and
$Z$-boson propagators in the SM. As can be seen from 
the RPV Lagrangian given in Eqs. \ref{gen_Lag1} and \ref{gen_Lag2},
the sneutrinos contribute through an $s$-channel exchange only 
when both $\lambda$ and $\lambda'$ couplings are simultaneously non-zero. 
The amplitude due to the $t$-channel exchange of squarks 
is non-zero when only $\lambda'$ couplings are non-vanishing.
It is straightforward to write down the amplitudes for the above
process, and the RPV contributions are given by
\beq\label{M12}
\ba{lll}
{\mathcal M}_1 &=& -i \lambda'_{j33} \lambda_{j11}\left[\bar{u}(p_1) {\mathcal P}_L v(p_2)\right]
\left[\bar{v}(k_2,s_2){\mathcal P}_R u(k_1,s_1)\right]/(s - M_{\tilde{\nu}_j}^2+
i \Gamma M_{\tilde{\nu}_j}), \\
{\mathcal M}_2 &=& -i \lambda_{1j3}^{'2} \left[\bar{u}(p_1){\mathcal P}_L u(k_1,s_1)\right]
\left[\bar{v}(k_2,s_2) {\mathcal P}_R v(p_2)\right]/(t -
M_{\tilde{u}_j}^2),
\ea
\eeq
where ${\mathcal P}_L,{\mathcal P}_R$ are the left and right chirality
projection matrices.

The total amplitude may be written as
\begin{equation}
{\mathcal M} = {\mathcal M}_{\gamma} + {\mathcal M}_Z + {\mathcal M}_1 +
{\mathcal M}_2,
\end{equation}
where $ {\mathcal M}_{\gamma,Z}$ are the amplitudes for
$s$-channel $\gamma$ and $Z$ exchanges, and ${\mathcal M}_{1,2}$ given
in Eq. \ref{M12} are the amplitudes for the RPV diagrams in
Fig. \ref{fig1}. The detailed formulae for all interference terms are collected in the
Appendix.  
We find in the interference terms, dependent on the RPV couplings
quadratically rather than quartically,  the characteristic  $\sin \theta
\cos \phi$ / $\sin \theta \sin \phi$ dependence of the term linear
in $P_T^{e^-/e^+}$.
It is also interesting to note that the term ${\mathcal M}_2^*{\mathcal M}_1$
corresponding to the interference of the $s$-channel and $t$-channel
RPV amplitudes is not symmetric in the $e^-$ and $e^+$ polarizations.
This peculiar structure is the result of the chiral nature of the RPV 
couplings. Moreover, the term gives a non-zero contribution only when the
electron beam has transverse polarization, and the positron beam has
longitudinal (or no) polarization. The chiral nature of the RPV
couplings is also reflected in the fact that the term vanishes for
vanishing final state fermion mass $M_f$.

The above treatment can be very easily generalized to the case
of $t\bar{t}$ production too. In that case, the $M_f^2/s$ suppression
encountered in the case of the $b\bar b$ final state would 
be considerably reduced. One must however note that we do not have
any $s$-channel contribution for up-type quarks in the final state, a
fact which is obvious
from the structure of the RPV Lagrangian given in
Eq. \ref{gen_Lag1} and Eq. \ref{gen_Lag2}.

We now focus on the contributions to the $b\bar{b}$ final state,
coming from the s-/t-channel scalar exchange due to the
RPV couplings  and compare them with the
SM expectations. For simplicity, we have considered the cases of
only sneutrino exchange in the $s$ channel or only squark exchange in the
$t$ channel.  This is sufficient, as normally one considers the case where only
the relevant RPV couplings are non-zero.
For studying sneutrino exchange we consider only one non-vanishing
(or dominant) combination of the
$\lambda$ and $\lambda'$ couplings, while for the
squark exchange contributions we consider only one non-vanishing (or dominant)
$\lambda'$-coupling. Simultaneous presence of more than one coupling could
potentially lead to flavour-changing neutral currents and hence is subject
to rather stringent constraints. Of course, needless to say that
in our numerical studies presented in the following subsections,
we choose values of the couplings consistent with these constraints.

We have performed our numerical studies in the context of an ILC operating
with a center-of-mass energy ($\sqrt{s}$) of 500 GeV and the choice 
of transverse polarization for the colliding beams is $(+0.8,+0.6)$. The
final state fermions satisfy the kinematic cuts:
\bi
\item The minimum transverse momenta $p_T^f$ of the fermions should be 20 GeV.
\item The fermions should not be close to the beam pipe and must respect the
angular cut
of $10^\circ < \theta^f < 170^\circ $.
\ei
\subsubsection{Sneutrino exchange in the s-channel}

In this section we discuss the case where the only non-zero RPV
contribution to the process $e^+e^-\to b\bar{b}$  is
via sneutrino exchanges in the $s$ channel ( Fig. \ref{fig1}a).
This means that the relevant
non-zero RPV coupling combination would be
$\lambda_{j11} \lambda'_{j33}$, where $j$ corresponds to the sneutrino flavor
($\snu_j$). Due to the antisymmetry property of the $\lambda_{ijk}$ couplings
in its first two indices, we know that only $j=2,3$ for the sneutrino flavor
can contribute in the $s$-channel. This further ensures that the relevant
$\lambda'$ couplings will be $\lambda'_{233}$ or $\lambda'_{333}$. The
$\lambda'$ couplings that can contribute in the $t$-channel squark exchange
must have the form $\lambda'_{1k3}$ where $k$ determines the squark flavor.
Thus assuming one non-vanishing $\lambda'_{j33}$ and the others to be
zero corresponds to the situation
that when the sneutrino diagram contributes, the squark exchange diagram
\begin{figure}[htb]
\begin{center}
\includegraphics[height=2.5in,width=2.9in]{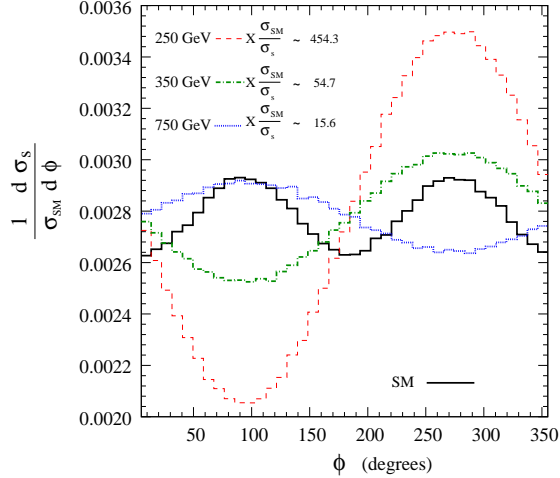}
\caption{\sl\small The normalized differential cross-section for the
R-parity violating contribution
as a function of the azimuthal angle for different values of sneutrino
mass. The coupling constants are chosen for each sneutrino mass to
saturate the experimental bounds, as discussed in the text.
Also shown in solid lines is the SM expectation.}
\label{fig2}
\end{center}
\end{figure}
would not. Since we would like to
restrict ourselves to a single non-zero $\lambda$ coupling at a time,
we have chosen
\beq\label{lamlamp}
\lambda_{211}\lambda'_{233} \lsim 7.2 \times 10^{-4}
                         \left(\frac{M_{\snu_j}}{100~GeV}\right)^2,
\eeq
which is consistent with limits estimated from LEP for the above process.

\begin{figure}[htb] 
\begin{center}
\includegraphics[height=2.4in,width=2.6in]{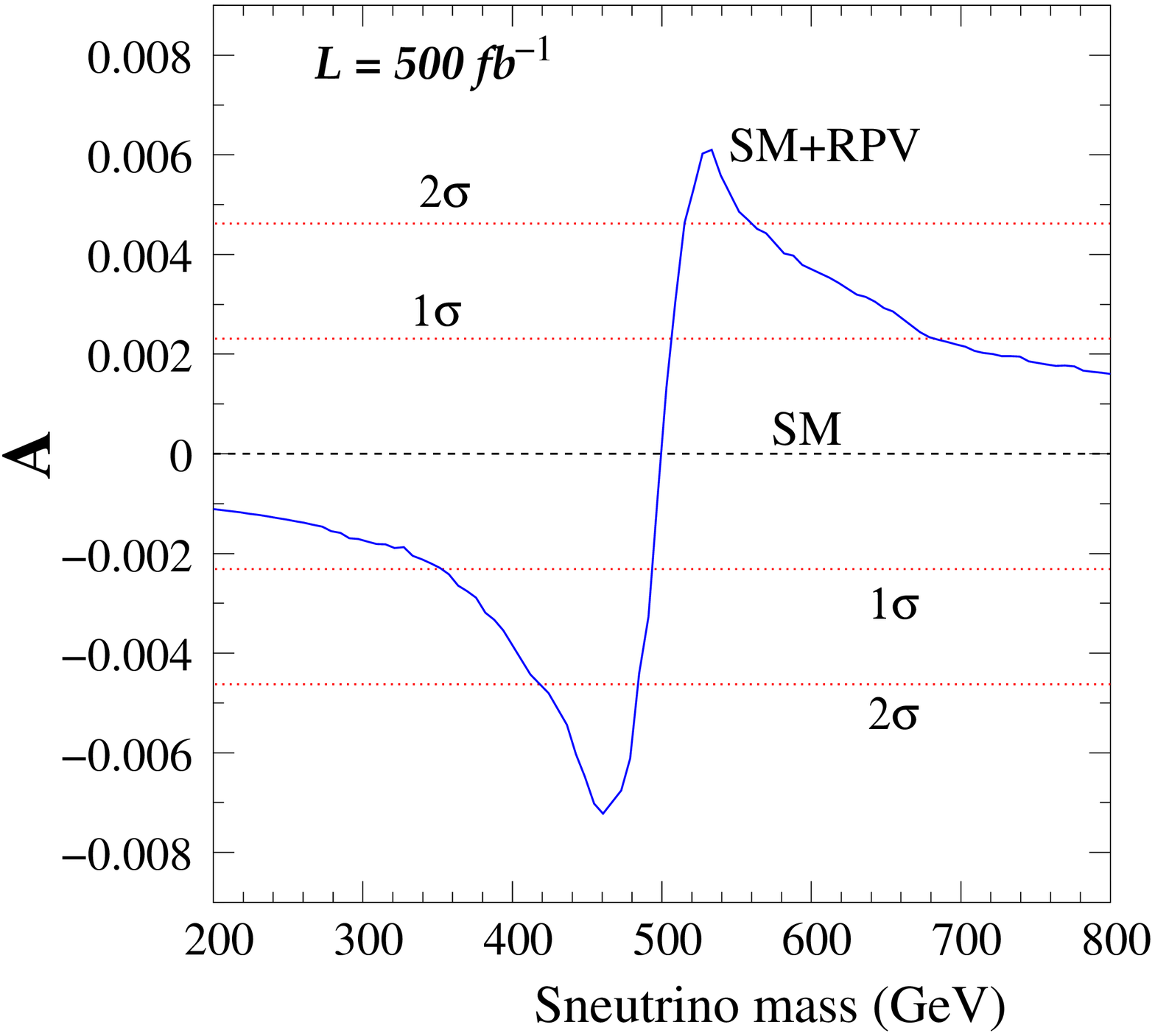}
\includegraphics[height=2.4in,width=2.6in]{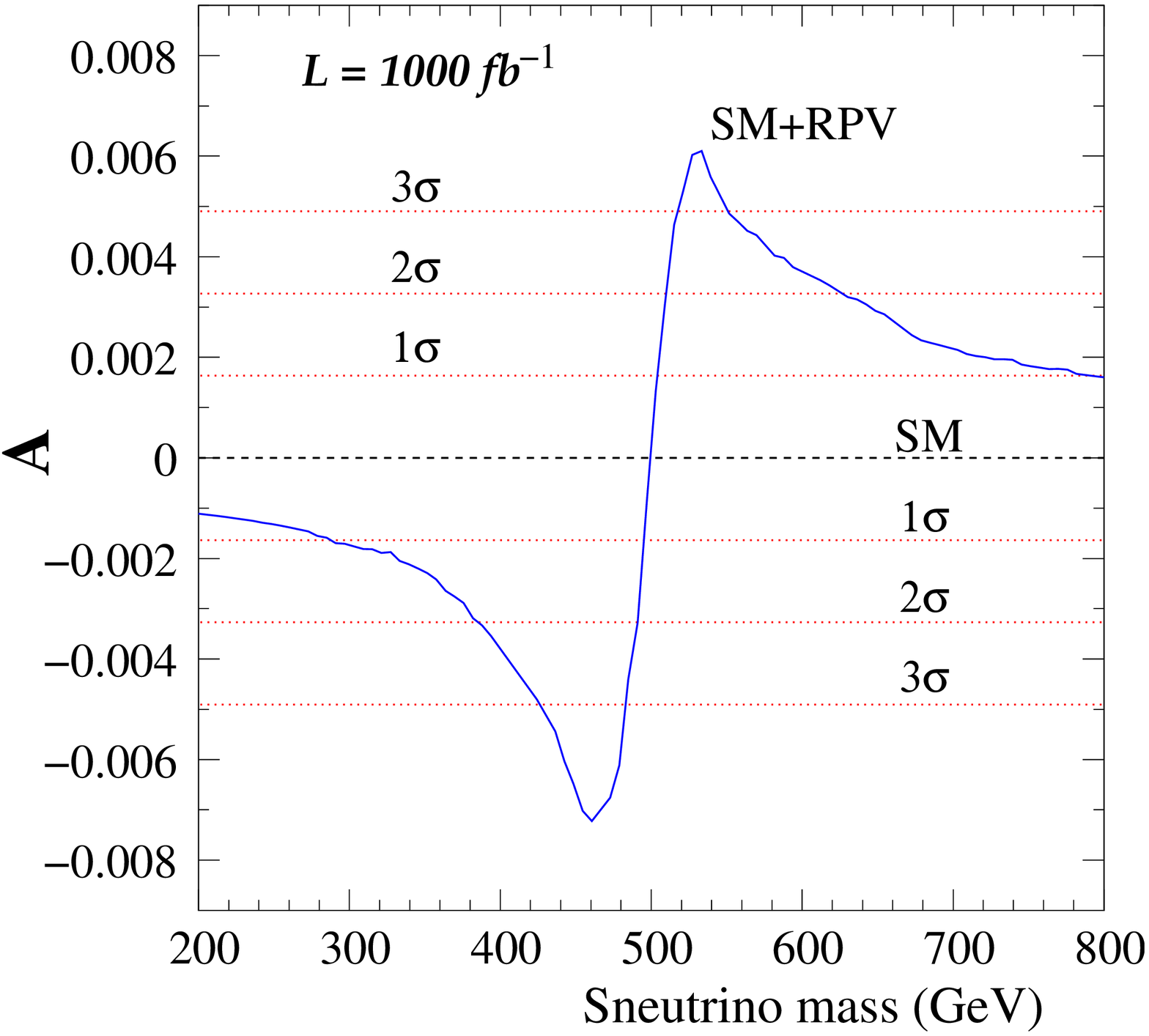}
\caption{\sl\small The asymmetry $A$ as defined by Eq. \ref{asnu}
for the signal as a function of the sneutrino mass for integrated luminosities
$L=500$, 1000~fb$^{-1}$ and for the maximum value of the product of RPV 
couplings $\lambda_{211}\lambda'_{233}$ for that sneutrino mass. 
Also shown are the SM expectation and discovery limits at $1\sigma$, 
$2\sigma$ and $3\sigma$ levels.}
\label{fig3}
\end{center}
\end{figure}

In Fig. \ref{fig2} we show the normalized differential cross section
dependence on the azimuthal angle $\phi$ for the SM as well as for the excess
over the SM for different values of the sneutrino mass, with the
combinations of $\lambda$ and $\lambda'$ chosen to saturate the
experimental bounds of Eq. \ref{lamlamp}. Thus, we have used the values
$\lambda_{211}\lambda'_{233} = 0.0045, 0.0088, 0.0405$ for $M_{\snu_j} =
250, 350, 750$ GeV, respectively.
The azimuthal angle is defined with respect to the direction of $e^-$ as
the $z$ axis and the transverse polarization direction of $e^-$ as the $x$
axis.
It is clear from Fig. \ref{fig2} that the distribution for the SM is symmetric 
about $\phi=\pi$. It can also be checked that pure sneutrino exchange also
produces a symmetric distribution. However, there is a marked asymmetry about
$\phi=\pi$ for the interference between the SM and
the RPV contributions. We define an asymmetry  which isolates
the new physics contribution, given by
\beq
A = \frac{\sigma (0 < \phi \le \pi) - \sigma (\pi < \phi \le 2\pi)}
               {\sigma (0 < \phi \le 2\pi)}
\label{asnu}
\eeq
A quick look at Fig. \ref{fig2} shows that this asymmetry vanishes for the
SM. We note that this azimuthal dependence for the $s$-channel
exchange is proportional to the mass of the final state fermion, which
in this case is the mass of the $b$ quark. Thus we would not have expected
any azimuthal dependence if the final state had massless fermions.
Note also that a sneutrino of mass 500 GeV then would be
produced at the peak of a resonance.
In this case the asymmetry identically vanishes as the  dominant
contribution comes from the direct term of sneutrino exchange. This is also
highlighted in Fig. \ref{fig3}, where we show the asymmetry $A$ as a function of
the sneutrino mass, for two different integrated luminosities, 
viz., $L=500$ and 1000~fb$^{-1}$. We allow the coupling product to
scale to the maximum value as allowed for that particular mass of the
sneutrino, given by Eq. \ref{lamlamp}.

The figure also shows corresponding to each luminosity the asymmetry
values needed to differentiate the RPV model from SM at 1
$\sigma$ and 2 $\sigma$ levels, and also at the 3 $\sigma$ level, in
case of Fig. \ref{fig3}(b).


We find that the asymmetry is quite sensitive to the mass of the
sneutrino. It has the expected structure of a resonant term interfering
with a non-resonant one. Thus it peaks for sneutrino masses which are very 
close to the center-of-mass energy $\sqrt{s}$ of the machine, and goes 
through a zero at $\sqrt{s}$ as seen in Fig. \ref{fig3}(a) and (b).
In the asymmetry we have defined, and which is shown in Fig. \ref{fig3},
SM contributions appearing in the denominator are large, and hence there is
a huge suppression of the asymmetry.

\subsubsection{Squark exchange in the t-channel}

In this section we discuss the case where the only non-zero RPV
couplings contribute to $e^+e^-\to b\bar{b}$
via $t$-channel squark exchange. In this case the
non-zero RPV couplings would have the form
$\lambda'_{1j3}$ where $j$ corresponds to the generation index of the
exchanged squark, and
the sneutrino exchange term vanishes as all $\lambda$ couplings are
assumed to be zero. We have restricted our choice to
RPV $\lambda'_{1j3}$ coupling which satisfies \cite{rplimits}
\beq\label{lamp}
\lambda'_{1j3}\le 0.02\left
     (\frac{M_{\widetilde{q}_{j}}}{100~GeV}\right).
\eeq
\begin{figure}[htb]
\begin{center}
\includegraphics[height=2.5in,width=2.5in]{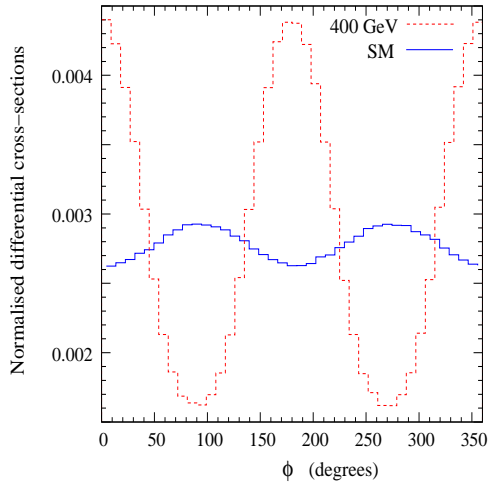}
\caption{\sl\small The normalized differential cross-section
for the R-parity violating contribution
as a function of the azimuthal angle for $t$-channel exchange of a squark of
mass 400 GeV. Also shown in solid lines is the SM expectation.}
\label{fig5}
\end{center}
\end{figure}
In Fig. \ref{fig5} we show by broken lines the dependence on the azimuthal 
angle $\phi$ of the normalized differential cross section for the excess over 
the SM, where
the squark exchanged in the $t$-channel has a mass of 400 GeV. The solid
lines represent the SM expectation, identical to what was shown in
Fig. \ref{fig2}. It is clear that the contributions from the new physics to
the azimuthal distribution are quite different from that of the SM. Also,
\begin{figure}[htb]
\begin{center}
\includegraphics[height=2.4in,width=2.6in]{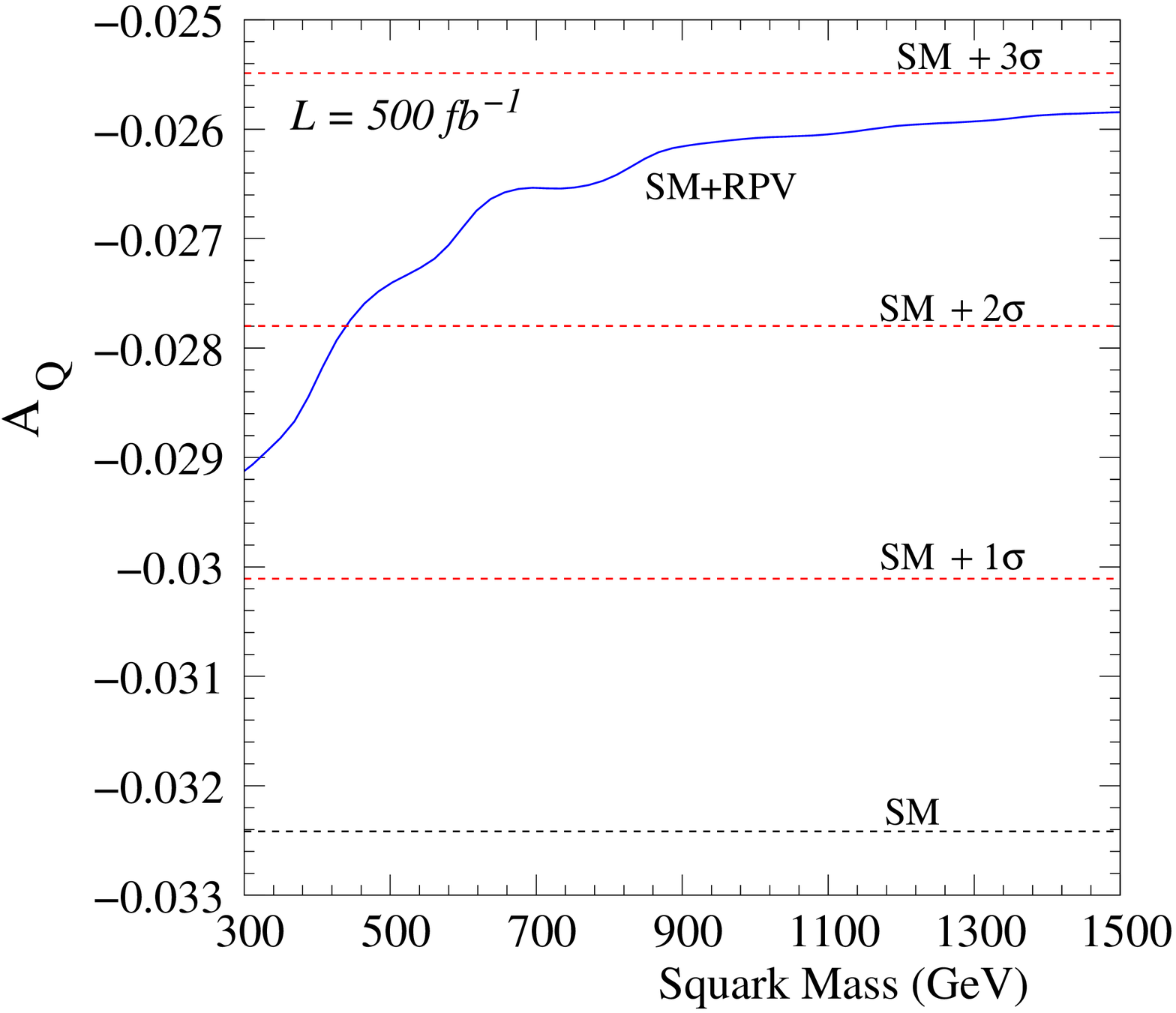}
\includegraphics[height=2.4in,width=2.6in]{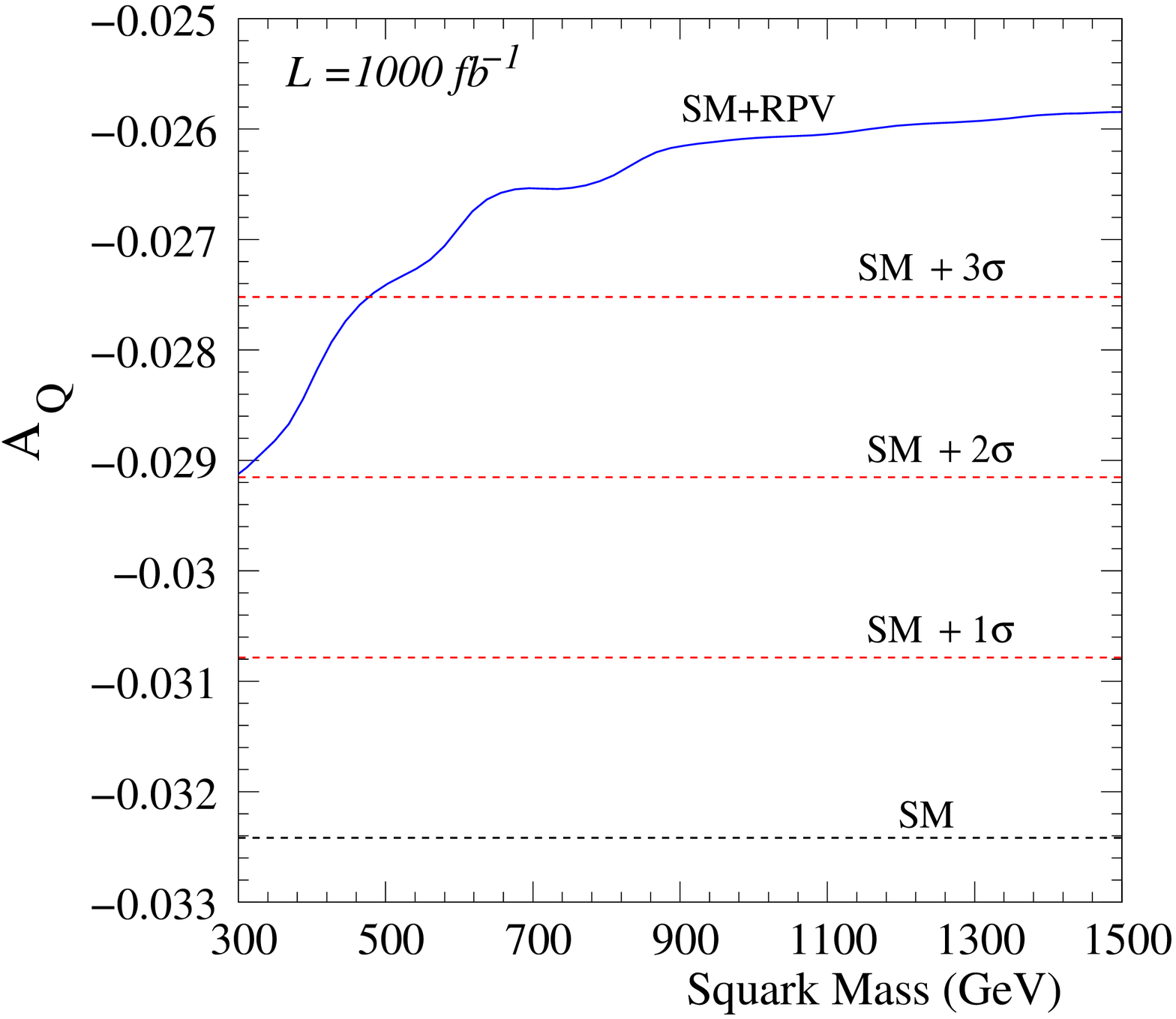}
\caption{\sl\small The asymmetry $A_Q$ in the azimuthal distribution
for $t$-channel exchange of squark as a function of the squark mass exchanged
for the maximum allowed value of $\lambda$ for that squark mass
for integrated luminosities $L=500,1000$~fb$^{-1}$.
Also shown are the SM expectation and discovery limits at $1\sigma$, $2\sigma$ and $3\sigma$
levels.}
\label{fig6}
\end{center}
\end{figure}
the previously defined asymmetry vanishes identically for both the SM as well
as the new physics contribution. So, to highlight the RPV
contribution, we define a
new asymmetry in the azimuthal angle,
\beq
A_{Q} = \frac{\sigma(0<\phi\le\frac{\pi}{4})
         -\sigma(\frac{\pi}{4}<\phi\le \frac{3\pi}{4})
         +\sigma(\frac{3\pi}{4}<\phi\le\frac{5\pi}{4})
         -\sigma(\frac{5\pi}{4}<\phi\le\frac{7\pi}{4})
         +\sigma(\frac{7\pi}{4} < \phi \le 2\pi)}
               {\sigma (0 < \phi \le 2\pi)}
\label{asquark}
\eeq
In Fig. \ref{fig6} we plot against squark mass the asymmetry
$A_Q$ for both SM and the total signal
(SM+RPV in the figure) corresponding to two different integrated
luminosities, viz., $L=500$ and 1000~fb$^{-1}$.
The RPV coupling
is allowed to scale to its maximum permissible value corresponding
to the mass of the squark.
The figure also shows corresponding to each luminosity the asymmetry
values needed to differentiate the RPV model from SM at 1
$\sigma$, 2$\sigma$ and 3 $\sigma$ levels, where use has been made of the relation
\beq\label{sigmalevel}
\vert A_Q - A_{\mathrm SM}\vert = n
		\frac{\sqrt{1-A_{\mathrm SM}^2}}{\sqrt{L\sigma_{SM}}},
\eeq
for the deviation of the asymmetry $A_Q$ from the SM asymmetry $A_{\mathrm SM}$
by $n \sigma$ for an SM cross section $\sigma_{SM}$ and integrated
luminosity $L$.
It can be seen that the defined asymmetry can easily differentiate the RPV
contributions from the SM one. With the high luminosity expected to
be available at the ILC, it
will be able to differentiate even for very small values of RPV
couplings, or equivalently, low squark masses,
at the $3\sigma$ level.

\section {Conclusions} \label{sec-4}
We have thus investigated the special role played by transverse polarization
in probing the RPV couplings at an  $e^+e^-$ collider. We illustrate this 
with  the process $e^+ e^- \rightarrow b \bar b$. Transverse polarization
in fact allows us to construct azimuthal asymmetries  which can probe
contribution coming from the interference terms, dependent 
on the  RPV couplings only quadratically. These asymmetries
help us isolate the contribution coming from RPV couplings and thus
offer interesting possibilities for probing them. We find,
using the currently allowed maximum values of the $\lambda$ couplings,
that these help us probe squark masses over a wide range
much beyond the energy available at the collider. Alternatively, the
increased  sensitivity at lower squark masses possible for  higher luminosity
indicates reach to lower values of the RPV couplings beyond the
current limits.

We have also  compared the possible sensitivity  using  total cross section
for the process with longitudinal polarization, with the one obtainable using
azimuthal asymmetries with transversely polarized beams. 
We find that the sensitivity of the total cross section 
in case of $s$-channel sneutrino exchange is larger by a factor 
varying from about 10 to about 500. 
This can be attributed to the severe suppression coming from 
the $b$ mass in the azimuthal asymmetries.  Having observed such an
excess the next challenge is to identify the new physics responsible for it.
In principle, the polar-angle distribution could be used to 
discriminate RPV theory from the SM and other 
theories like extra $Z$ models, as was done for example in 
\cite{Rizzo:1998vf}. However, it was seen in \cite{Rizzo:1998vf} that 
 the sensitivity of polar distribution for a leptonic final state is rather 
low,  with or without longitudinal polarization.  If we were to  use the same 
strategy as that of ref. \cite{Rizzo:1998vf} we see, using the results therein,
 after correcting them for the values of $s$, luminosity as well as
a different  ($b\bar b$)  final state, that the ratio of RPV
couplings to sneutrino mass for which the polar distribution can
discriminate the model is above that allowed by present experimental
limits.

In case of $t$-channel squark exchange, the sensitivity of the total cross 
section is higher by a factor of order 5, compared to that from the
azimuthal asymmetries,  even though there is no suppression due to the $b$ 
mass. Again, extrapolating the results of  Ref.  \cite{Rizzo:1998vf}, which 
considered $t$-channel sneutrino exchange for a leptonic final state 
to our case, we expect that the corresponding ratio of coupling to squark mass 
is ruled out.

The azimuthal asymmetries then show that in the region still allowed by
the current data, the effects would be  beyond $2\sigma$ or $3\sigma$
fluctuations of the SM expectations and hence can be probed. Thus they have 
a higher sensitivity to these couplings, i.e.,
the change in the polar-angle distributions
expected with the present limits on the couplings will be not be
beyond fluctuations of the SM  whereas the azimuthal asymmetries would
be.

In principle, if the $b$-mass
suppression were not there, then transverse polarization could have
provided even  higher sensitivity.
Nevertheless,
azimuthal asymmetries we consider here have a greater reach for RPV
couplings than polar distributions.
Moreover, the azimuthal asymmetries we use for the $s$-channel sneutrino
case are vanishing for chirality-conserving couplings in theories like
extra $Z$ theories, or extra-dimensional theories with massive spin-1 or
massive graviton exchange in the $s$ channel.  Hence their presence
would clearly discriminate the RPV theory with its  chirality-violating
couplings from the rest.

\section {Acknowledgements} \label{acknow}
R.G. wishes to acknowledge support from the Department of Science and 
Technology, India under Grant No. SR/S2/JCB-64/2007, under the J.C. Bose
Fellowship scheme. S.K.R. gratefully acknowledges support from the
Academy of Finland (Project No. 115032). 
\section{Appendix} \label{appendix}
 The amplitudes with all interference terms for the new physics signal, {\it i.e.} 
the RPV contributions are given explicitly as
\begin{eqnarray*}
 {\mathcal M}^*_1 {\mathcal M}_1 &=& \frac{{\lambda'}^2_{j33} \lambda^2_{j11}}{16}
                     \frac{1}{(s - M_{\snu_j}^2)^2
 + \Gamma^2 M_{\snu_j}^2} (1 + \PpL + \PeL + \PeL\PpL) (4 s^2 - 5sM_f^2) \\
{\mathcal M}^*_2 {\mathcal M}_1 &=& \frac{{\lambda'}^2_{1j3} {\lambda'}_{j33}
\lambda_{j11}}{8}
                    \frac{ s\sqrt{(s - M_f^2)}}{(s - M_{\snu_j}^2
 + i \Gamma M_{\snu_j}) (t - M_{\tilde{u}_j}^2)}
(1+\PpL)\PeT M_f \sin\theta (- \cos\phi + i \sin\phi) \\
{\mathcal M}^*_\gamma {\mathcal M}_1 &=& \frac{\pi Q_e Q_f {\lambda'}_{j33} \lambda_{j11}}{\alpha}
 \frac{1}{s (s - M_{\snu_j}^2+ i \Gamma M_{\snu_j})}
 4 s \sqrt{(s - M_f^2)} M_f \sin\theta \\
 &\times& \left[ \PeT(1+\PpL) (\cos\phi + i \sin\phi)+\PpT(1+\PeL)
(\cos\phi - i \sin\phi) \right] \\
{\mathcal M}^*_\gamma {\mathcal M}_2 &=& -\frac{\pi Q_e Q_f {\lambda'}^2_{1j3}}{\alpha}
            \frac{1}{s (t - M_{\tilde{u}_j}^2)} \\
   &\times&    \left[ (1 + \PpL - \PeL - \PeL\PpL)   (
         2 u^2 + 4 s u + 2 s^2 - 4 M_f^2 u - 2 M_f^2 s + 2 M_f^4) \right.\\
 && \left.  - \PeT \PpT (2 u^2 + 2 s u - 4 M_f^2 u + 2 M_f^4
     + s \sin^2\theta \cos\phi (\cos\phi - i \sin\phi) (s - M_f^2)) \right ] \\
 {\mathcal M}^*_Z {\mathcal M}_1 &=& \frac{\pi {\lambda'}_{j33} \lambda_{j11}}{8 \alpha s_W^2 c_W^2}
         \frac{s \sqrt{(s - M_f^2)}M_f f_V ~\sin\theta}
              {(s-M_Z^2-i\Gamma_Z M_Z)(s-M_{\snu_j}^2+i\Gamma M_{\snu_j})} \\
  &\times& \left[ \PpT (1 + \PeL) (c_A+c_V) (\cos\phi - i \sin\phi)
          - \PeT (1 + \PpL) (c_A-c_V) (\cos\phi + i \sin\phi) \right] \\
 {\mathcal M}^*_Z {\mathcal M}_2 &=& \frac{\pi {\lambda'}^2_{1j3}}{16 \alpha s_W^2 c_W^2}
                  \frac{}{(s-M_Z^2-i\Gamma_Z M_Z) (t - M_{\tilde{u}_j}^2)} \\
 &\times& \left[\PeT \PpT  (c_A+c_V) (f_A+f_V)
   (2 u^2 + 2 s u - 4 M_f^2 u + 2 M_f^4 \right. \\
 && \left. + (s^2 - s M_f^2) \sin^2\theta \cos\phi ( \cos\phi - i \sin\phi))
 + 2 (1 - \PeL + \PpL - \PeL \PpL) (c_A -c_V) \right. \\
&& \left.  ( (f_A + f_V) (u^2 + 2s u + s^2 + M_f^4)
          - f_A M_f^2 (2 u + 3 s)
          - f_V M_f^2 (2 u + s)) \right ] \\
 {\mathcal M}^*_2 {\mathcal M}_2 &=& \frac{{\lambda'}^4_{1j3}}{4}
                     \frac{1}{(t - M_{\tilde{u}_j}^2)^2}
 (1 + \PpL - \PeL - \PeL \PpL)  (
          u^2 + 2 s u + s^2 - 2 M_f^2 u - 2 M_f^2 s + M_f^4).
\end{eqnarray*}
Note that these
contain the full dependence on both longitudinal and
transverse polarization. It is a trivial exercise to isolate amplitudes
for the individual cases, depending on the choice of beam polarization. In
our study we will be interested only in the case where the beams are
transversely polarized.
In the above expressions, $M_f$ stands for the fermion mass in the
final state, which in this case is the $b$-quark mass. For the couplings of
quarks and leptons to the $Z$ boson the following notation is used:
$$c_V = 2 T_3^e - 4 Q_e s_W^2,~ c_A= -2 T_3^e,$$ and
$$f_V = 2 T_3^f - 4 Q_f s_W^2,~ f_A= -2 T_3^f.$$

For completeness we also present
below the direct amplitudes in the SM, coming from the exchange of $\gamma$
and the $Z$ boson in the $s$-channel. Since the beams would be either
longitudinally or transversely polarized, we give the SM amplitudes for both
the cases separately.
\bi
\item {\it SM amplitudes with longitudinally polarized beams:}
\begin{eqnarray*}
\vert {\mathcal M}_{\g}\vert^2 &=& Q_e^2 Q_f^2 e^4 \frac{N_c^f (1 - P_L^{e^-} P_L^{e^+})}{s^2}
              \left[4u^2+4su+2s^2+4M_f^4-8uM_f^2\right]\\
{\mathcal M}_{\g}^* {\mathcal M}_Z &=& \frac{Q_e Q_f e^4}{16 s_W^2 c_W^2}
\frac{N_c^f \left[(1 - P_L^{e^-} P_L^{e^+})c_V f_V
+ (P_L^{e^-}-P_L^{e^+})c_A f_V\right]}
{s~(s-{\mathcal M}_Z^2+i\G_Z{\mathcal M}_Z)} \\
&\times& \left[(4u^2+4su+2s^2+4M_f^4-8uM_f^2)\right] \\
 &-& \frac{Q_e Q_f e^4}{16 s_W^2 c_W^2}
\frac{N_c^f \left[(1 - P_L^{e^-} P_L^{e^+})c_A f_A
+ (P_L^{e^-}-P_L^{e^+})c_V f_A\right]}
{s~(s-{\mathcal M}_Z^2+i\G_Z{\mathcal M}_Z)} \left[(2s^2+4su-4sM_f^2)\right] \\
\vert {\mathcal M}_{Z}\vert^2 &=&  \frac{e^4}{256 s_W^4 c_W^4}
\frac{N_c^f (f_V^2+f_A^2)\left[(1 - P_L^{e^-} P_L^{e^+})(c_V^2+c_A^2)
+2(P_L^{e^-}-P_L^{e^+})c_V c_A \right]}
{(s-{\mathcal M}_Z^2)^2+(\G_Z{\mathcal M}_Z)^2} \\
&& \times \left[(4u^2+4su+2s^2+4M_f^4-8uM_f^2)\right] \\
&-&  \frac{e^4}{256 s_W^4 c_W^4}
\frac{N_c^f \left[4(1 - P_L^{e^-} P_L^{e^+})\right]} {(s-{\mathcal M}_Z^2)^2+(\G_Z{\mathcal M}_Z)^2} \\
&\times& \left[c_V c_A f_V f_A (2s^2+4su-4sM_f^2) + 2 s M_f^2 f_A^2 (c_V^2+c_A^2)\right] \\
&-&  \frac{e^4}{256 s_W^4 c_W^4}
\frac{N_c^f \left[2(P_L^{e^-}- P_L^{e^+})\right]}{(s-{\mathcal M}_Z^2)^2+(\G_Z{\mathcal M}_Z)^2}
\left[f_V f_A (c_V^2+c_A^2) (2s^2+4su-4sM_f^2)\right]
\end{eqnarray*}
\item {\it SM amplitudes with transversely polarized beams:}
\begin{eqnarray*}
\vert{\mathcal M}_{\g}\vert^2 &=& \frac{Q_e^2 Q_f^2 e^4 N_c^f}{s^2}
\left [(1 - P_T^{e^-} P_T^{e^+}) (4u^2+4su-8uM_f^2+4M_f^4) \right. \\
&& \left. + (2 s^2 + P_T^{e^-} P_T^{e^+}
(2sM_f^2-2s^2) \sin^2\theta\cos^2\phi)\right]\\
{\mathcal M}_{\g}^* {\mathcal M}_Z &=& \frac{Q_e Q_f e^4}{16 s_W^2 c_W^2}
\frac{N_c^f} {s~(s-{\mathcal M}_Z^2+i\G_Z{\mathcal M}_Z)} \\
&\times& \left[(1 - P_T^{e^-} P_T^{e^+})c_V f_V(4u^2+4su+4M_f^4-8uM_f^2) - c_A f_A (2s^2 + 4su - 4s M_f^2) \right. \\
&& \left. + 2 c_V f_V s^2 + c_V f_V P_T^{e^-} P_T^{e^+} (2sM_f^2 - 2s^2) \sin^2\theta\cos^2\phi \right. \\
&& \left.  + i c_A f_V P_T^{e^-} P_T^{e^+} (2sM_f^2 - 2s^2) \sin^2\theta\sin\phi\cos\phi \right] \\
\vert {\mathcal M}_Z\vert ^2 &=&  \frac{e^4}{256 s_W^4 c_W^4}
\frac{N_c^f} {(s-{\mathcal M}_Z^2)^2+(\G_Z{\mathcal M}_Z)^2} \\
&\times& \left[(1 - P_T^{e^-} P_T^{e^+})(c_V^2+c_A^2)(f_V^2+f_A^2)(4u^2+4su
+4M_f^4-8uM_f^2)\right. \\
&&+ \left. \left\{ f_A^2(c_V^2+c_A^2)(2s^2-8sM_f^2)
+ 2 s^2 f_V^2 (c_V^2+c_A^2)
- 8 c_V c_A f_V f_A (s^2 + 2su  \right.\right. \\
&& \left.\left. -2 s M_f^2) + P_T^{e^-} P_T^{e^+}(c_V^2+c_A^2)(f_V^2+f_A^2)(2s^2-2 s M_f^2)
\sin^2\theta\cos^2\phi \right\}\right].
\end{eqnarray*}
\ei
We have followed a notation similar to that mentioned above for the
fermion couplings.
\def\pr#1,#2,#3 { {\em Phys.~Rev.}        ~{\bf #1},  #2 (#3) }
\def\prd#1,#2,#3{ {\em Phys.~Rev.}        ~{\bf D#1}, #2 (#3) }
\def\prl#1,#2,#3{ {\em Phys.~Rev.~Lett.}  ~{\bf #1},  #2 (#3) }
\def\plb#1,#2,#3{ {\em Phys.~Lett.}       ~{\bf B#1}, #2 (#3) }
\def\npb#1,#2,#3{ {\em Nucl.~Phys.}       ~{\bf B#1}, #2 (#3) }
\def\prp#1,#2,#3{ {\em Phys.~Rept.}       ~{\bf #1},  #2 (#3) }
\def\zpc#1,#2,#3{ {\em Z.~Phys.}          ~{\bf C#1}, #2 (#3) }
\def\epj#1,#2,#3{ {\em Eur.~Phys.~J.}     ~{\bf C#1}, #2 (#3) }
\def\mpl#1,#2,#3{ {\em Mod.~Phys.~Lett.}  ~{\bf A#1}, #2 (#3) }
\def\ijmp#1,#2,#3{{\em Int.~J.~Mod.~Phys.}~{\bf A#1}, #2 (#3) }
\def\ptp#1,#2,#3{ {\em Prog.~Theor.~Phys.}~{\bf #1},  #2 (#3) }
\newcommand{\rmp}[3]{Rev. Mod.  Phys. {\bf #1} #2 (#3)}             %
\newcommand{\rpp}[3]{Rep. Prog. Phys. {\bf #1} #2 (#3)}             %
\newcommand{\npbps}[3]{Nucl. Phys. B (Proc. Suppl.)
           {\bf #1} #2 (#3)} %
\newcommand{\sci}[3]{Science {\bf #1} #2 (#3)}                 %
 \newcommand{\apj}[3]{ Astrophys. J.\/ {\bf #1} #2 (#3)}       %
\newcommand{\jhep}[2]{{Jour. High Energy Phys.\/} {\bf #1} (#2) }%
\newcommand{\astropp}[3]{Astropart. Phys. {\bf #1} #2 (#3)}            %
\newcommand{\ib}[3]{{ ibid.\/} {\bf #1} #2 (#3)}                    %
\newcommand{\nat}[3]{Nature (London) {\bf #1} #2 (#3)}         %
 \newcommand{\app}[3]{{ Acta Phys. Polon.   B\/}{\bf #1} #2 (#3)}%
\newcommand{\nuovocim}[3]{Nuovo Cim. {\bf C#1} #2 (#3)}         %
\newcommand{\yadfiz}[4]{Yad. Fiz. {\bf #1} #2 (#3);             %
Sov. J. Nucl.  Phys. {\bf #1} #3 (#4)]}               %
\newcommand{\jetp}[6]{{Zh. Eksp. Teor. Fiz.\/} {\bf #1} (#3) #2;
           {JETP } {\bf #4} (#6) #5}%
\newcommand{\philt}[3]{Phil. Trans. Roy. Soc. London A {\bf #1} #2
        (#3)}                                                          %
\newcommand{\hepph}[1]{hep--ph/#1}           %
\newcommand{\hepex}[1]{hep--ex/#1}           %
\newcommand{\astro}[1]{(astro--ph/#1)}         %
\newcommand{\etal}{{\em et al.}}



\begin{thebibliography}{99}

\bibitem{Fayet:1977yc}
  P.~Fayet,
  Phys.\ Lett.\  B {\bf 69}, 489 (1977);
  G.~R.~Farrar and P.~Fayet,
  Phys.\ Lett.\  B {\bf 76}, 575 (1978).

\bibitem{Ibanez:1992pr}
L.~E.~Ibanez and G.~G.~Ross,
Nucl.\ Phys.\ {\bf B368} (1992) 3.


\bibitem{books}
See e.g.~M. Drees, R.M. Godbole and P. Roy,  {\it Theory and
phenomenology of sparticles}, World Scientific, 2005;
H.~Baer and X.~Tata,
{\it ``Weak scale Supersymmetry: From superfields to scattering events,''}
{\it  Cambridge, UK: Univ. Pr. (2006) }.


\bibitem{Dreiner:1992vm}
  H.~K.~Dreiner and G.~G.~Ross,
  Nucl.\ Phys.\  B {\bf 410}, 188 (1993)
  [arXiv:hep-ph/9207221].

\bibitem{decay}
E.A. Baltz and P. Gondolo, Phys. Rev. {\bf D57}  2969 (1998).
H.K. Dreiner, P. Richardson and M.H. Seymour, JHEP {\bf 0004}, 008 (2000);
F. Borzumati, R.M. Godbole, J.L. Kneur and F. Takayama, JHEP {\bf 07}, 037
(2002).

\bibitem{dreinergrab}
H.~K.~Dreiner and S.~Grab,
  arXiv:0811.0200 [hep-ph]

\bibitem{rpvcol}
See, for example,
R.M. Godbole, P. Roy and X. Tata, Nucl. Phys. {\bf
B401} 67 (1993);
D.K. Ghosh, R.M. Godbole and S. Raychaudhuri,
Z. Phys. {\bf C75} (1997) 375;
 D.~Choudhury, S.~K.~Rai and S.~Raychaudhuri,
  Phys.\ Rev.\  D {\bf 71}, 095009 (2005)
  A.~Datta and S.~Poddar,
  Phys.\ Rev.\  D {\bf 75}, 075013 (2007)

\bibitem{nosusy}
  D.~Choudhury and S.~Raychaudhuri,
  Phys.\ Rev.\  D {\bf 56}, 1778 (1997)
  [arXiv:hep-ph/9703369];
  F.~Abe {\it et al.}  [CDF Collaboration],
  Phys.\ Rev.\ Lett.\  {\bf 83}, 2133 (1999)
  [arXiv:hep-ex/9908063];
  B.~Abbott {\it et al.}  [D0 Collaboration],
  Phys.\ Rev.\ Lett.\  {\bf 83}, 4476 (1999)
  [arXiv:hep-ex/9907019];
  A.~A.~Affolder {\it et al.}  [CDF Collaboration],
  Phys.\ Rev.\ Lett.\  {\bf 88}, 041801 (2002)
  [arXiv:hep-ex/0106001].

\bibitem{HERA}
C.~Adloff \etal\ (H1 Collaboration), \zpc74,191,{1997};
J.~Breitweg \etal\ (ZEUS Collaboration), \zpc74,207,{1997}.

\bibitem{Barger:1989rk}
V. Barger, G. F. Giudice and T. Han,
\prd40,2987,{1989};
%
  G.~Bhattacharyya and D.~Choudhury,
  Mod.\ Phys.\ Lett.\  A {\bf 10}, 1699 (1995)
  [arXiv:hep-ph/9503263].

\bibitem{rplimits}
For a summary of these limits, see for example, \\
G.~Bhattacharyya, hep-ph/9709395;
B.C.~Allanach, A.~Dedes and H.K.~Dreiner, \prd60,075014,{1999};
M.~Chemtob, Prog. Part. Nucl. Phys.~{\bf 54} 71 (2005);
  R.~Barbier {\it et al.},
  Phys.\ Rept.\  {\bf 420}, 1 (2005).

\bibitem{Bhattacharyya:1994yc}
G. Bhattacharyya, D. Choudhury and K. Sridhar,
\plb349,118,{1995};
D.~Choudhury,  R.M.~Godbole and G.~Polesello,  JHEP~{\bf 0208}, 004
(2002);
%
J. Kalinowski, R. R\"{u}ckl, H. Spiesberger and P. M. Zerwas,
\plb414,297,{1997};
%
J. L. Hewett and T. G. Rizzo,
\prd56,5709,{1997};
%
D. K. Ghosh, S. Raychaudhuri and K. Sridhar,
\plb396,177,{1997}.

\bibitem{Hikasa:1999wy}
E.L.~Berger, B.W.~Harris and Z.~Sullivan, \prd63,115001,{2001};
K. Hikasa, J. M. Yang and B. Young, \prd60,114041,{1999};
D.~Choudhury, \plb346,291,{1995};
A.~Datta, \prd65,054019,{2002}.

\bibitem{CCQR}M.~Carena, D.~Choudhury, S.~Lola and C.~Quigg,
        \prd58,095003,{1998};
         M.~Carena, D.~Choudhury, C.~Quigg and S.~Raychaudhuri,
        \prd62,095010,{2000}.
%
D.~Choudhury and S.~Raychaudhuri, hep-ph/9807373.

\bibitem{ILC}
E. Accomando {\it et al.}, Phys. Rept. 299 (1998) 1
[arXiv:hep-ph/9705442];
J. Aguilar-Saavedra {\it et al.},  hep-ph/0106315; T. Abe {\it  et
al.},  hep-ex/0106055-58; K. Abe {\it et al},   hep-ph/0109166,
J.~Brau {\it et al.}, SLAC-R-857;
A.~Djouadi, J.~Lykken, K.~Monig, Y.~Okada, M.~J.~Oreglia and S.~Yamashita,
  arXiv:0709.1893 [hep-ph].


\bibitem{gudi}
  G.~A.~Moortgat-Pick {\it et al.},
  Phys.\ Rept.\  {\bf 460}, 131 (2008)
  [arXiv:hep-ph/0507011].

\bibitem{newphys}
  T.~G.~Rizzo,
  JHEP {\bf 0302}, 008 (2003)
  [arXiv:hep-ph/0211374];
  JHEP {\bf 0308}, 051 (2003)
  [arXiv:hep-ph/0306283];
  B.~Ananthanarayan, S.~D.~Rindani, R.~K.~Singh and A.~Bartl,
  Phys.\ Lett.\  B {\bf 593}, 95 (2004)
  [Erratum-ibid.\  B {\bf 608}, 274 (2005)]
  [arXiv:hep-ph/0404106];
  B.~Ananthanarayan and S.~D.~Rindani,
  Phys.\ Lett.\  B {\bf 606}, 107 (2005)
  [arXiv:hep-ph/0410084];
  JHEP {\bf 0510}, 077 (2005)
  [arXiv:hep-ph/0507037];
  K.~Rao and S.~D.~Rindani,
  Phys.\ Lett.\  B {\bf 642}, 85 (2006)
  [arXiv:hep-ph/0605298];
  Phys.\ Rev.\  D {\bf 77}, 015009 (2008)
  [arXiv:0709.2591 [hep-ph]];
  S.~D.~Rindani and P.~Sharma,
  Phys.\ Rev.\  D {\bf 79}, 075007 (2009)
  [arXiv:0901.2821 [hep-ph]];
  S.~S.~Biswal and R.~M.~Godbole,
  arXiv:0906.5471 [hep-ph];
  K.~Huitu and S.~K.~Rai,
  Phys.\ Rev.\  D {\bf 77}, 035015 (2008)
  [arXiv:0711.4754 [hep-ph]].

\bibitem{lepto}
  S.~D.~Rindani,
  Phys.\ Lett.\  B {\bf 602}, 97 (2004)
  [arXiv:hep-ph/0408083].

\bibitem{hikasa} K.-i.~Hikasa,
  Phys.\ Rev.\  D {\bf 33}, 3203 (1986).
\bibitem{ttbar}
  B.~Ananthanarayan and S.~D.~Rindani,
  Phys.\ Rev.\  D {\bf 70}, 036005 (2004)
  [arXiv:hep-ph/0309260].

\bibitem{Rizzo:1998vf}
  T.~G.~Rizzo,
  Phys.\ Rev.\  D {\bf 59}, 113004 (1999)
  [arXiv:hep-ph/9811440].

\end{thebibliography}
\end{document}